%% file: paper-finalish-1.tex
\documentclass{acm_proc_article-sp}
\pdfoutput=1
\usepackage{mathtools,amssymb,placeins}
\usepackage[normalem]{ulem}

\usepackage{graphicx,multirow}

\newcommand{\papertitle}{A Method for Reducing the Severity of Epidemics by Allocating Vaccines According to Centrality}
\title{\papertitle}
\numberofauthors{3}
\author{
\alignauthor 
Krzysztof Drewniak\\
\email{kdrewniak@utexas.edu}
\alignauthor Joseph Helsing\\
\email{JosephHelsing@my.unt.edu}
\alignauthor Armin R Mikler\\
\email{mikler@cs.unt.edu}}

\begin{document}
\maketitle{}

\begin{abstract}
One long-standing question in epidemiological research is how best to allocate limited amounts of vaccine or similar preventative measures in order to minimize the severity of an epidemic.
Much of the literature on the problem of vaccine allocation has focused on influenza epidemics and used mathematical models of epidemic spread to determine the effectiveness of proposed methods.
Our work applies computational models of epidemics to the problem of geographically allocating a limited number of vaccines within several Texas counties.
We developed a graph-based, stochastic model for epidemics that is based on the SEIR model, and tested vaccine allocation methods based on multiple centrality measures.
This approach provides an alternative method for addressing the vaccine allocation problem, which can be combined with more conventional approaches to yield more effective epidemic suppression strategies.
We found that allocation methods based on in-degree and inverse betweenness centralities tended to be the most effective at containing epidemics.
\end{abstract}
\section{Introduction}
One long-standing question in epidemiological research is how best to allocate limited amounts of vaccine or similar preventative measures in order to minimize the severity of an epidemic.
Discovering a way to reduce the severity of epidemics would benefit society by not only preventing loss of life, but also reducing the overall disease burden of those epidemics.
These burdens, or societal costs, can be seen in reduced economic productivity due to widespread sickness, or increased strain on health infrastructure.

Much of the literature on the problem of vaccine allocation has focused on influenza epidemics and has used mathematical models of epidemic spread to determine the effectiveness of proposed methods \cite{Matrajt2010,Medlock2009,Mylius2008,Tuite2010,Wallinga2010}.
Most investigations of vaccine allocation performed their analyses on geographically large scales and focused on determining which sub-populations should be prioritized for vaccination.
Since previous research focuses on large-scale, subpopulation-based models, our work explores alternative approaches to the question of vaccine allocation.

Much work has been devoted to the development of computational models for disease simulation.
Mikler \emph{et. al.} have developed multiple stochastic models of disease spread in a population, which use the standard SEIR(S) model \cite{Mikler2009, Reyes-Silveyra2011}.
Some of these models, such as their Global Stochastic Cellular Automata model \cite{Mikler2005}, can be adapted to the problem of vaccine allocation.
Simulation techniques relevant to our work have been developed by Indrakanti \cite{Indrakanti2012}, who implemented a SEIR-based framework for simulating epidemic spread within a county.

Our work applies computational models of epidemics to the problem of geographically allocating a limited number of vaccines within several Texas counties.
We developed a graph-based, stochastic model for epidemics that is based on the SEIR model and the work of Mikler \emph{et. al.} \cite{Mikler2005}.
Various centrality measures have been proposed over the years \cite{Nieminen1973,Bonacich1972,Bonacich2007,Freeman1977,Brandes2001}, mainly in the field of social network analysis, which provide means for determining which nodes in a graph are the most important.
Our model was then used to investigate various centrality-based vaccine allocation strategies, in which vaccines are allocated to various census blocks within the county in order of their centrality measure score.
This approach provides an alternative method for addressing the vaccine allocation problem, which can be combined with more conventional approaches to yield more effective epidemic suppression strategies.
In this paper, we present the results of several experiments with centrality-based vaccine allocation strategies.
These experiments were performed on graphs constructed from the census data of several Texas counties, and have yielded results that indicate directions for further study.

\section{Background}
\subsection{Vaccine Allocation}
Several authors have previously addressed the problem of vaccine allocation.
Medlock and Galvani have investigated the question of which groups should be prioritized for influenza vaccination throughout the United States in the event of a shortage \cite{Medlock2009}.
They developed a model, parameterized with real-world data, for analyzing various vaccination strategies under multiple effectiveness criteria.
Medlock and Galvani found that optimal epidemic control could be achieved by targeting adults between the ages of 30 and 39 and children, and that current CDC vaccine allocation recommendations were suboptimal under all considered measures.
Tuite \emph{et. al.} used similar methods to investigate vaccine allocation strategies in Canada during a H1N1 epidemic \cite{Tuite2010}.
They found that targeting high-risk groups, such as people with chronic medical conditions, and then targeting age groups, such as children, that are the most likely to develop complications after influenza infection will cause the greatest reduction in the number of deaths and serious illnesses caused by an influenza epidemic.

Similar work was performed by Matrajt and Longini \cite{Matrajt2010}.
In their work, Matrajt and Longini attempted to determine whether the optimal vaccine allocation strategy varies with the state of the epidemic.
They developed an SIR-based model to analyze the spread of influenza among and between several groups of people in idealized countries, and used criteria similar to those of Medlock and Galvani to ascertain the effectiveness of their methods.
The authors determined that the demographic structure of the population in an area significantly affects the optimal allocation strategy.
They also found that, at the beginning of an epidemic, those groups that are the most likely to transmit influenza should be targeted immediately, while the groups that are most vulnerable should be targeted after a point before, but near, the peak of the epidemic.
The work of Mylius \emph{et. al} modeled influenza epidemics and found that individuals that are most at risk for complications should be targeted if additional vaccines become available during an epidemic \cite{Mylius2008}.
Mylius \emph{et. al.} also found that schoolchildren should be prioritized for vaccination at the beginning of an epidemic, as they are heavily involved in disease transmission.
Their conclusions bear a close resemblance to those of Matrajt and Longini.

The problem of large-scale vaccine allocations were addressed by \cite{Matrajt2010,Medlock2009,Mylius2008,Tuite2010}.
They used different models to reach their conclusions, but used similar metrics, such as the reduction in the predicted number of deaths from a disease after the vaccination program, to assess the significance of their results.
None of these works addressed the question of vaccine allocation at the level of counties or other similar geographic divisions, nor did they address the vaccine allocation problem from a geographic perspective.

Another approach to the vaccine allocation problem, which we extend in our work, can be found in the work of Johnson \cite{Johnson2010}.
Johnson's approach, unlike those of many other researchers, focuses on vaccine allocation in the small scale.
Johnson generated synthetic social network graphs and used various measures of centrality to determine which individuals within those graphs should be vaccinated.
Her results showed that the optimal vaccination strategy depends on the structure of the social network it is applied to.
Our work adapts Johnson's methods to the problem of allocating vaccinations within a county.

\subsection{Centrality and its Applications}
\label{sec:centrality-app}
The notion of centrality has been used in various fields, most notably in the study of social networks.
Multiple centrality measures exist, all of which allow the vertices of a graph to be ranked in order of their importance.
Out-Degree and in-degree centralities were first defined by Nieminen \cite{Nieminen1973}.
In those measures, the centrality $c$ of a node $n$ in a graph $G = (V,E)$, where $V$ is the set of nodes in the graph, $u$ and $v$ are nodes in V, $E$ is the set of edges, and $|V|$ is the number of nodes is
\begin{equation}
c = \frac{d_n}{|V| - 1}
\end{equation}
where $d_n$ is the out-degree or in-degree of $n$.

Eigenvector centrality, first proposed by Bonacich \cite{Bonacich1972}, of a graph $G = (V,E)$ can be calculated by representing $G$ as an adjacency matrix $A$, where
\begin{equation}
  A_{uv} = 
  \begin{cases}
    1 & \text{if } (u,v) \in E \\
    0 & \text{if } (u,v) \not\in E
  \end{cases}
\end{equation}
Then, the eigenvector centrality $x$ is defined as 
\begin{equation}
Ax = \lambda x
\end{equation}
where $\lambda$ is the largest eigenvalue of $A$.
Eigenvector centrality is useful because it gives a higher centrality score to a high-degree node that is connected to other high-degree nodes than to a high-degree node connected to low-degree nodes \cite{Bonacich2007}.

Another important measure of centrality is betweenness centrality.
Betweenness centrality, first described by Freeman \cite{Freeman1977}, is defined as follows, for a node $n$ in a graph $G = (V,E)$, where $d_{st}(n)$ is the number of shortest paths from $s \in V$ to $t \in V$ that include $n$, and $d_{st}$ is the number of shortest paths from $s$ to $t$ where the shortest path between two nodes is the path where the sum of the weights of its edges is minimized
\begin{equation}
\sum_{s \not = n \not = t \in V} \frac{d_{st}(n)}{d_{st}}
\end{equation}
Betweenness centrality gives higher centrality scores to nodes that are most likely to be involved in the transmission of information throughout a graph \cite{Brandes2001}.

Centrality has been applied to multiple epidemiological questions.
Rothenberg \emph{et. al.} applied various centrality measures to a social network graph generated from CDC data on the spread of HIV in Colorado Springs \cite{Rothenberg1995}.
Rothenberg analyzed the collected data under multiple centrality measures and found that all of the measures identified all but one of the HIV cases as non-central.
Their work also noted several differences in response patterns to the CDC questionnaire between people with high centrality and people with low centrality under all measures, which were closely correlated.

Similar work was also performed by Christley \emph{et. al.} \cite{Christley2005}.
Christley generated synthetic social networks and performed SIR-based simulations in order to study the applicability of centrality to the problem of identifying individuals at high risk of HIV infection.
They found that degree centrality performed no worse than other centrality measures, such as betweenness, but noted that the results might not be valid for larger datasets.

Centrality has also played a key role in Johnson's work, which was discussed previously \cite{Johnson2010}.
Johnson investigated how vaccinating those individuals identified as central in a social network would affect the spread of disease in that network.
Johnson found that vaccination strategies based on centrality measures were an effective means of mitigating outbreaks. 
Related work was performed by Rustam, who applied centrality to the spread of viruses and worms in computer networks \cite{Rustam2006}.
He found that nodes with high centrality scores have a large amount of influence on the spread of a virus, especially when those nodes are rated central by multiple measures.

\subsection{Computational Simulation of Epidemics}
Several methods for simulating epidemics computationally are based on the SEIR model of epidemics \cite{Mikler2009,Reyes-Silveyra2011,Mikler2005,Indrakanti2012}.
In the SEIR model, the population is divided into four classes: susceptible individuals, who can become infected; exposed or latent individuals, who have been infected but are not capable of spreading the infection; infectious individuals, who can spread the disease to susceptible individuals; and recovered individuals, who can no longer be infected.
Mikler \emph{et. al.} proposed a SEIR-based model that uses a stochastic cellular automaton \cite{Mikler2005}.
The concepts used in their model are applicable to a wide range of epidemic simulations.

Agent-based and metapopulation models are the two dominant methods for computational epidemic simulation \cite{Ajelli2010}.
In agent-based models, each individual within the population is simulated.
Such models, which are often implemented using cellular automata \cite{Mikler2005,Mikler2009,Fu2003}, are useful because they provide insights into the progression of the disease.
However, they are impractical when large populations need to be simulated due to space required for information about each agent in the simulation.
Metapopulation models, on the other hand, break the population into subpopulations and then simulate the interactions between and within the subpopulations \cite{Indrakanti2012,Venkatachalam2006}.
Such models sacrifice some of the precision offered by agent-based modeling in exchange for the ability to simulate large populations.
Our work uses a metapopulation-based model, where each census block constitutes a subpopulation.
This helps to overcome the excessive computational resources that would be required to simulate hundreds of thousands of individuals.

A closely related model that we have adapted for our own work has been developed by Indrakanti \cite{Indrakanti2012}.
Indrakanti developed a county-level epidemic simulator, which uses census blocks as the base unit of simulation.
The model allows for contacts to be generated between any two census blocks, and uses an interaction coefficient to determine the likelihood of contact.
The model was used to conduct several experiments which found that epidemics with higher infectivity (likelihood of spread during a contact) reached their peak percentage of infected individuals earlier.
It also discovered that the distribution of population between census blocks had a significant effect on the spread of disease.

\section{Methods}
\subsection{Representing Counties}
The United States Census Bureau provides geographic data on US counties. 
They subdivide counties into census blocks at the finest granularity, which are typically bounded by roads, streets, or water features.
Being that they may be varying sizes and shapes, rural census blocks are often significantly larger than urban ones.
Additionally, the population of a census block may vary greatly as there are many cencus-blocks that contain zero population, i.e. farmland, greenspace, or bodies of water, and others that contain high-desntiy residential structures, i.e. appartment buildings, that may have several hundred people \cite{cblockdef}.
This geographic data, along with the population of each census block, was obtained for multiple Texas counties, specifically Rockwall, Hays, and Denton County, and was stored in a PostGIS database.
The centroid of each census block, which is the average of the points defining the census block, was precomputed and stored within the database.

After the census block data was obtained, several methods for representing a county as a graph with census block nodes were investigated.
Rockwall County was used for these investigations due to its small size, allowing centrality measures to be computed quickly.
All of these representations were based on various methods of constructing a graph $G = (V,E)$ to represent a county, where $V$, the set of nodes, was the set of populated census blocks in the county.
Our first approach to representing counties as a graph, which was ineffective, was a representation where $E = \{(u,v) \mid u, v \in V, u \neq v, \textit{ST\_Touches}(u,v)\}$.
The \emph{ST\_Touches()} function, which determines whether two geographic entities are adjacent, was provided by PostGIS.
This representation used undirected edges.
Another of our early attempts was an approach where $E = \{(u,v) \mid u, v \in V, u \neq v, \textit{ST\_Distance}(\textit{Centroid}(u),v) < r\}$ for various constant values of $r$, where the \emph{ST\_Distance()} function was used to compute the shortest distance between the centroid of $u$ and the block $v$.
The \emph{Centroid()} function retrieves the centroid of a block from the database.
A variation of this approach where the $r$ value for each node $u \in V$ was lowered by multiplying it by 
\begin{equation}
1 - \frac{\textit{Population}(u)}{1000}
\end{equation}
on the assumption that the population of highly-populated blocks was more likely to interact within the block than that of sparsely-populated blocks was also investigated.
All of these approaches were ineffective and gave unrealistic results, such as the identification of a large rural block with a population of 5 as the most central block within Rockwall County.
Therefore, they were rejected in favor of a representation based on weighted edges.

The results of our investigations led us to use the following method to represent US counties in our experiments.
In our final model, the county was represented as a directed graph $G = (V,E)$, where $V$ was the set of populated census blocks.
The set of edges $E$ was equal to $\displaystyle \{(u,v,W_{u \rightarrow v} \mid u \in V, v \in V, u \neq v\}$, where $\displaystyle W_{u \rightarrow v}$ was the weight of the edge from $u$ to $v$.
The weights were computed by the following formula, which was adapted from the work of Mikler \emph{et. al.}\cite{Mikler2005}: {\small
\begin{equation}
W_{u \rightarrow v} = \frac{\textit{Population}(u) \times \textit{Population}(v)}{10000 \times \textit{ST\_Distance}(\textit{Centroid}(u), v) + 0.00001}
\label{eqn:weights}
\end{equation}}
This definition assigns higher weights to blocks that are more likely to interact, a fact that is used in our simulator.

The multiplication by $1/10000$ was included to offset the behavior of \emph{ST\_Distance()}, which returns distance as real value less than 1.
If the distances were left unscaled, the division could have produced weights outside of the range that can be accurately represented by real values in a computer.
The addition of $0.00001$ was used to address cases where the centroid of block $u$ is located within block $v$, which ordinarily results in a distance of $0$, from producing an error.
Once these weights were computed, the graphs were represented using the NetworkX library \cite{Hagberg2008-networkx}.

\subsection{Centrality Measures}
We modified the centrality measures described in section \ref{sec:centrality-app} to work with our representation.
Code from \cite{Hagberg2008-networkx} was used as a basis for our work.
Out-degree and in-degree centralities were trivially adapted; the out-degree and in-degree of a node were redefined to be the sum of the weights of the out-edges or in-edges, respectively, instead of a count of those edges.
The implementation of eigenvector centrality in NetworkX already allowed for weighted edges, so it was used without modification.
Since that implementation of eigenvector centrality used only the out-edges for its computations, a variation of the measure, eigenvector-in centrality, was also tested where the in-edges were used instead.

Betweenness centrality was the most heavily modified measure.
Since betweenness centrality relies on shortest paths, the reciprocals of all edge weights were taken before centrality computation, so that the shortest path between nodes $s$ and $t$ is the one that the disease is most likely to spread across.
Afterwards, the betweenness centrality scores were computed by a variation of Brandes's algorithm \cite{Brandes2001}, where Dijkstra's algorithm was applied to each node in parallel as opposed to sequentially. 
This modification was found to have no effect on the results of the centrality computation.
The centrality measure described above was reported as inverse betweenness centrality.

Finally, we defined random centrality, which is not a centrality measure, for use as a baseline.
In random centrality, each node is assigned a random real value in $[0,1)$, which is used as its centrality score.
The results of all centrality calculations except random centrality, which was re-computed before each set of simulations, were saved to avoid expensive recomputation when multiple experiments were performed on the same data.

\subsection{Graph-Based Stochastic Epidemic Simulation}
\begin{figure}
  \centering
  \includegraphics[scale=0.5]{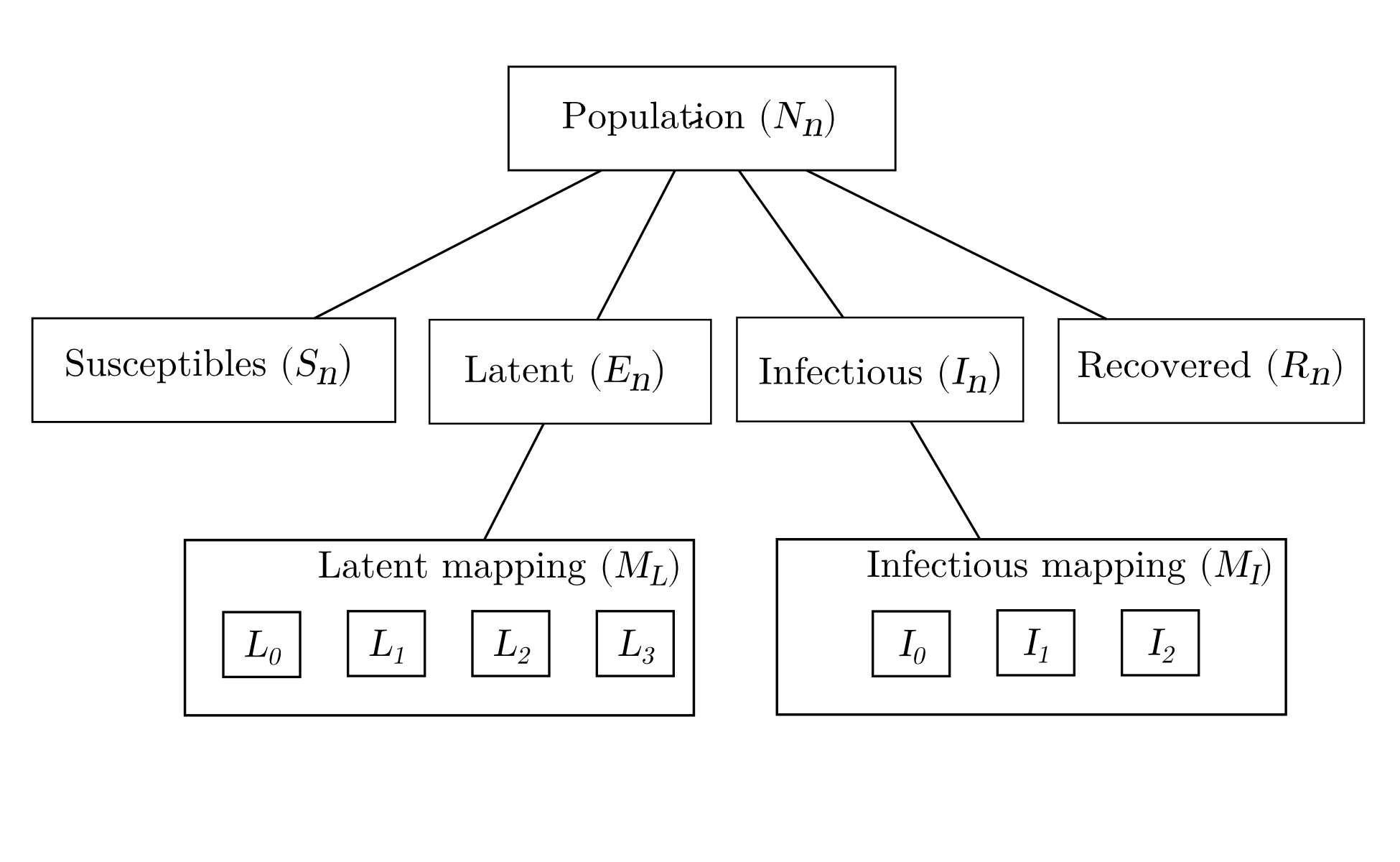}
  \caption{An illustration of the grouping of census blocks that have been split by the simulator.}
  \label{fig:data-model}
\end{figure}

A stochastic epidemic simulator based on the SEIR model and the work of \cite{Indrakanti2012} was developed for use with our representation of counties.
The simulation began by constructing the graph $G = (V,E)$ as described above, and computing all centrality measures that were to be tested.
Each node $n \in V$ was labeled with the following attributes: $|N_n|$, the total population of the block, $|S_n| = |N_n|$, the number of susceptibles, and $|E_n|, |I_n|, |R_n| = 0$, where $|E_n|$ is the number of latent (exposed) individuals in the block, $|I_n|$ the number of infectious, and $|R_n|$ the number of recovered individuals.
Throughout the simulation, $|S_n + E_n + I_n + R_n| = |N_n|$.
In addition to these counts, each node was labeled with a mapping $\displaystyle M_{E,n} = \{ 0 \mapsto E_{0,n}, 1 \mapsto E_{1,n} \ldots T_{lp} \mapsto E_{T_{lp},n}\}$, where $T_{lp}$ is the latent period and $E_{m,n}$ is the number of people in block $n$ with $0 \leq m \leq  T_{lp}$ days remaining until they transition to the next simulation state.
In this mapping, $\displaystyle \sum_{m=0}^{T_{lp}} E_{m,n} = |E_n|$.
A similar mapping was also produced for the infectious period, $\displaystyle M_{I,n} = \{ 0 \mapsto I_{0,n} \ldots T_{ip} \mapsto I_{T_{ip},n}$, where $T_{ip}$ is the infectious period.
For this mapping, $0 \leq m \leq T_{ip}$, similarly to $M_{E,n}$.
These attributes are illustrated in Figure \ref{fig:data-model}.

For the purposes of experimentation, 20 simulation sets were performed in parallel, with the random number generator re-seeded for each set.
Each set consisted of multiple simulations with varying values of available vaccine and centrality measures.
The simulations were performed on a server with four six-core Intel Xeon E7540 processors and 256 GB of RAM.

At the beginning of each set, a proportion $p$ of the census blocks were chosen randomly for initial infection, which remained constant throughout all simulations in the set.
The percentage of the population infected initially was varied between experiments.
Vaccines were distributed to the entire population of each chosen census until the total supply of vaccines was exhausted.
Census blocks were selected for vaccination in a decreasing order based on their centrality.

The centrality measure used to allocate vaccines was varied between experiments so that the measures could be compared for effectiveness.
The set of population (blocks) that was initially targeted for infection was held constant as the centrality measure and percent of population that was able to receive vaccination were varied.
This allowed us to more accurately determine which centrality measure or measures were most effective under varying conditions.
In experiments involving the prevention scenario, vaccines were distributed before the initial infection, while they were distributed six simulated days after initial infection in the intervention scenario.
The value of six days was chosen because at the beginning of day six the initial cases would have transitioned to the recovered state.
After the initial infections, which were recorded as day zero and placed those affected into $\displaystyle E_{T_{E},n}$, the simulation was executed until no latent or infectious individuals remained in the population.

During each day, after any scheduled vaccine distribution, each block's $M_E$ and $M_I$ is updated by setting
$M_E = \{ i \in M_E \mapsto M_E(i + 1), T_{lp} \mapsto 0 \}$, and similarly for $M_I$.
After these updates, the census blocks are iterated over.
For each block $n$ , $i$ contacts are simulated, where $i$ is $I_n$ multiplied by the contact rate, $20$ contacts/day, which was taken from \cite{Indrakanti2012}.
For each contact, a random real value in $[0,1)$ is generated.
If this number is greater than or equal to the mobility parameter, which was $0.99$, the contact takes place within $n$, otherwise, the contact takes place within a different census block.
The mobility parameter expresses the likelihood that an individual will contact someone who resides in a census block that the individual does not reside in.

\begin{figure}
  \centering
  \includegraphics[scale=0.45]{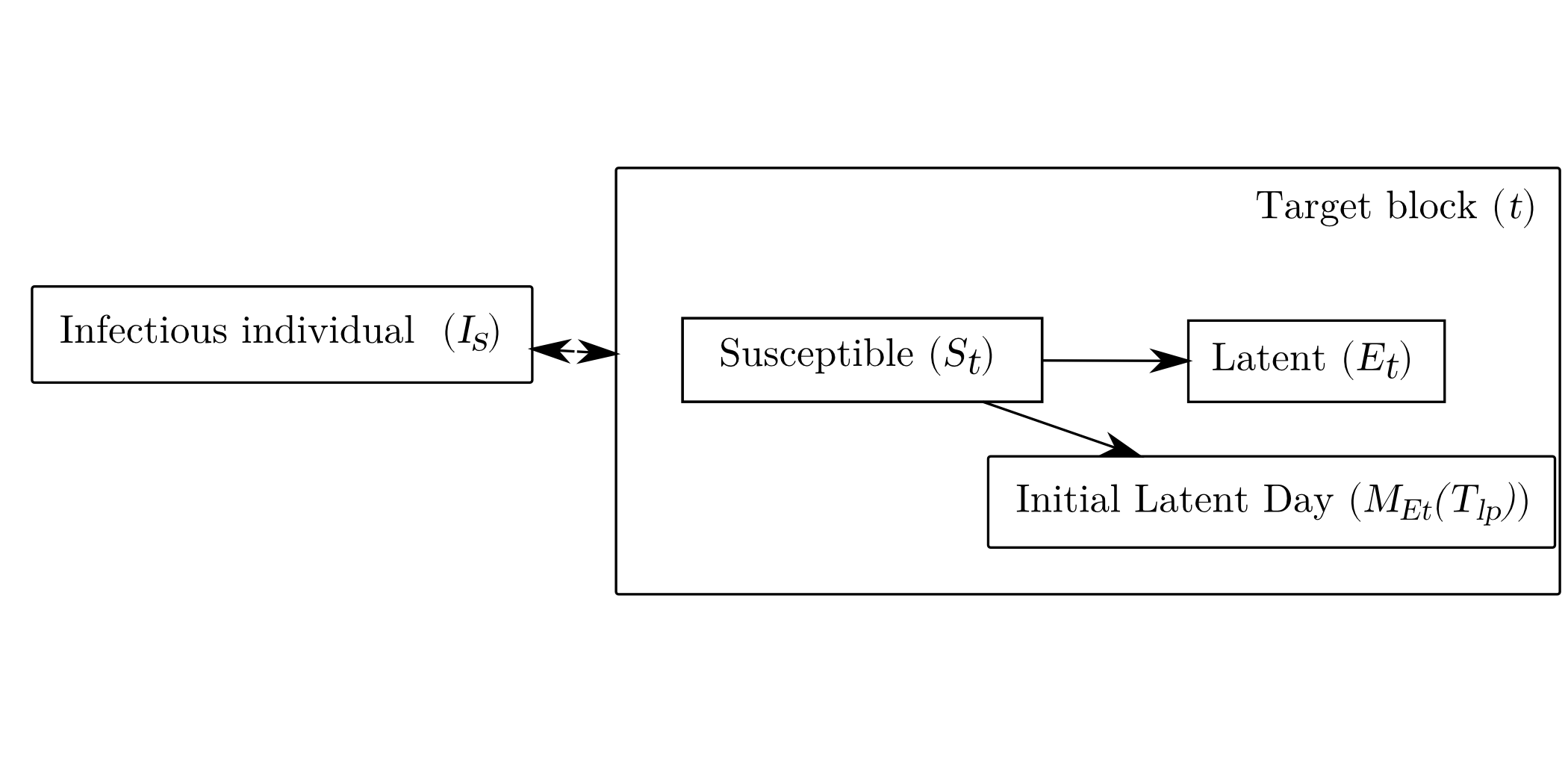}
  \caption{An illustration of the infection process used by the simulator.}
  \label{fig:infection-process}
\end{figure}

In the case of non-local contacts, i.e. contacts made between agents in separate blocks, the external block is chosen by the following method.
First, the weights (given by Equation \ref{eqn:weights}) of the out-edges of the originating block are normalized by dividing by the out-degree of the block and then sorted in descending order.
These values were precomputed once as an optimization.
A random target value in $[0,1)$ is generated, and the list of weights is summed element by element until the sum is greater than or equal to the target value.
The node $t$ whose associated weight causes the summation to terminate is chosen as the block for contact.
This process is illustrated by Figure \ref{fig:infection-process}.

Once the block that will be contacted is chosen, a disease-spreading contact occurs if a random real number in $[0,1)$ is less than the transmissibility parameter, which was $0.05$.
For non-local contacts, this test is performed before block selection to speed up the simulation significantly.
When the contact occurs, one person is, if possible, removed from $S_t$, the susceptible population of the target block $t$, and added to $E_t$, the latent population.
The new infection is also added to $M_{Et}(T_{lp})$, which ensures that the newly-infected individual remains in the latent period for $T_{lp}$ days before becoming infectious. 

After all the contacts are generated, the SEIR states are updated.
For each block $n \in V$, $M_{En}(0)$ people are removed from $E_n$, and added to $M_{In}(T_{ip})$ and $I_n$.
Similarly, $M_{In}(0)$ people are transferred from $I_n$ to $R_n$.
At the conclusion of the updates, the number of people throughout the county in each SEIR state and the percentage of the population in that state are reported.
In addition, the number of people who are infected, that is, either latent or infectious, is reported, along with the associated percentage.

\subsection{Experiments and Parameters}
In all of the experiments performed, the parameters listed in Table \ref{tab:params} were used.
These parameters were not chosen to reflect an extant disease, and were adapted from the work of Indrakanti \cite{Indrakanti2012}.
\begin{table}
  \centering
  \begin{tabular}{|l r|}
    \hline
    Contact rate & 20\\
    Transmissibility & 0.05\\
    Mobility & 0.99\\
    Latent period ($T_{lp}$) & 2 days\\
    Infectious period ($T_{ip}$) & 3 days\\
    Percentage of blocks to infect ($p$) & 1\%\\ \hline
  \end{tabular}
  \caption{Parameters kept constant throughout all experiments. These parameters serve to characterize the disease being simulated.}
  \label{tab:params}
\end{table}

Experiments were performed on Rockwall, Hays, and Denton counties using two epidemic scenarios for each county.
In the prevention scenario, vaccines were distributed before any infection took place, and 5 percent the population of the blocks that were selected for infection was initially infected.
In the intervention scenario 50 percent of the population of the infected blocks were infected initially, and at the beginning of the 6th day of the simulation vaccines were distributed. 
These scenarios were intended to simulate a naturally-occurring epidemic of a disease such as influenza or a random, mass-exposure bio-terror attack using perhaps smallpox, respectively.
Each combination of county and scenario received its own run of 20 simulation sets.

Within each simulation set, the following experiments were performed.
The initially infected blocks were held constant for all of these experiments within a set.
The amount of vaccine available was varied between 30 percent, 50 percent, 75 percent, and 90 percent of the population of the county.
Out-degree, in-degree, eigenvector, eigenvector-in, inverse betweenness, and random centralities were tested.

\section{Results}
In Tables \ref{tab:peaksp} and \ref{tab:peaksi}, we present the average peak infected percentage from our experiments.
This data was obtained by first finding the maximum percent infected for each simulation.
Then, the maximums of each of the 20 simulations for each experiment were arithmetically averaged.
The maximum percentage infected at any given time indicates the severity of the epidemic and, consequently, the strain that epidemic places on health resources, such as hospitals.
These averages, along with their corresponding standard deviations, are reported in Table \ref{tab:peaksp} for the prevention scenario, and Table \ref{tab:peaksi} for the intervention scenario.
These results have been rounded to four decimal places to avoid the appearance of false precision.

\begin{figure}
  \centering
  \includegraphics[scale=0.55]{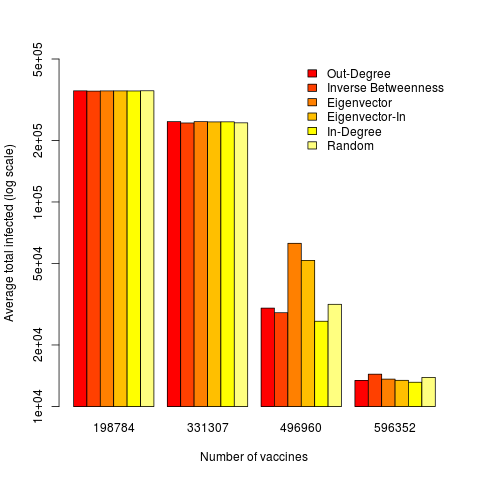}
  \caption{Average total number of infected individuals vs. number of vaccines for Denton County intervention scenario experiments, log scale.}
  \label{fig:rptotals}
\end{figure}

We also obtained the total number of infected individuals for each simulation execution.
These totals allow us to determine the severity of infections that do not result in outbreaks.
These results were averaged together on a per-experiment basis and plotted on a log scale.
The plot for an intervention scenario in Denton County is presented in Figure \ref{fig:rptotals}.

\begin{figure*}
  \centering
  \includegraphics[scale=0.4]{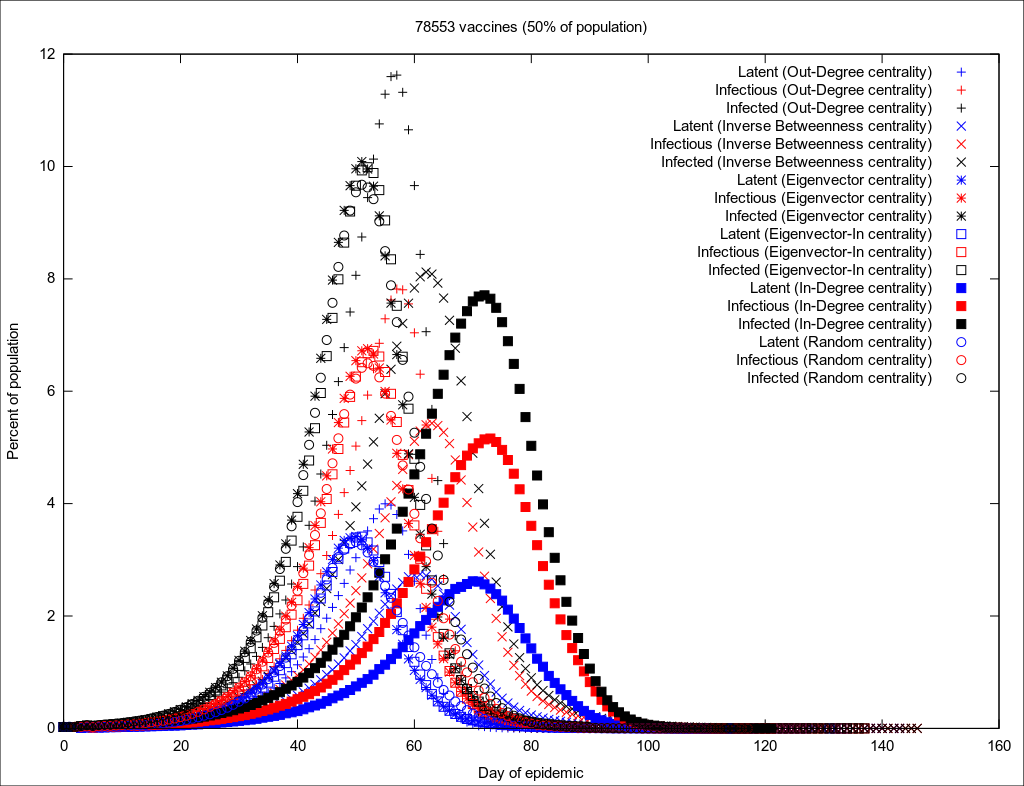}
  \caption{The average percentage of the population that was latent, infectious, or infected (either latent or infectious) v. time for Hays County prevention scenario (50 percent vaccination). Points with a value of 0 were omitted from the graph.}
  \label{fig:hinice}
\end{figure*}

We also plotted the average percentage of the population that was in each state during every day of the simulation for each experiment to better visualize the progress of the simulated epidemics.
Such a plot for a prevention scenario in Hays County at 50 percent vaccination is included as Figure \ref{fig:hinice}.

\begin{figure}
  \centering
  \includegraphics[scale=0.3]{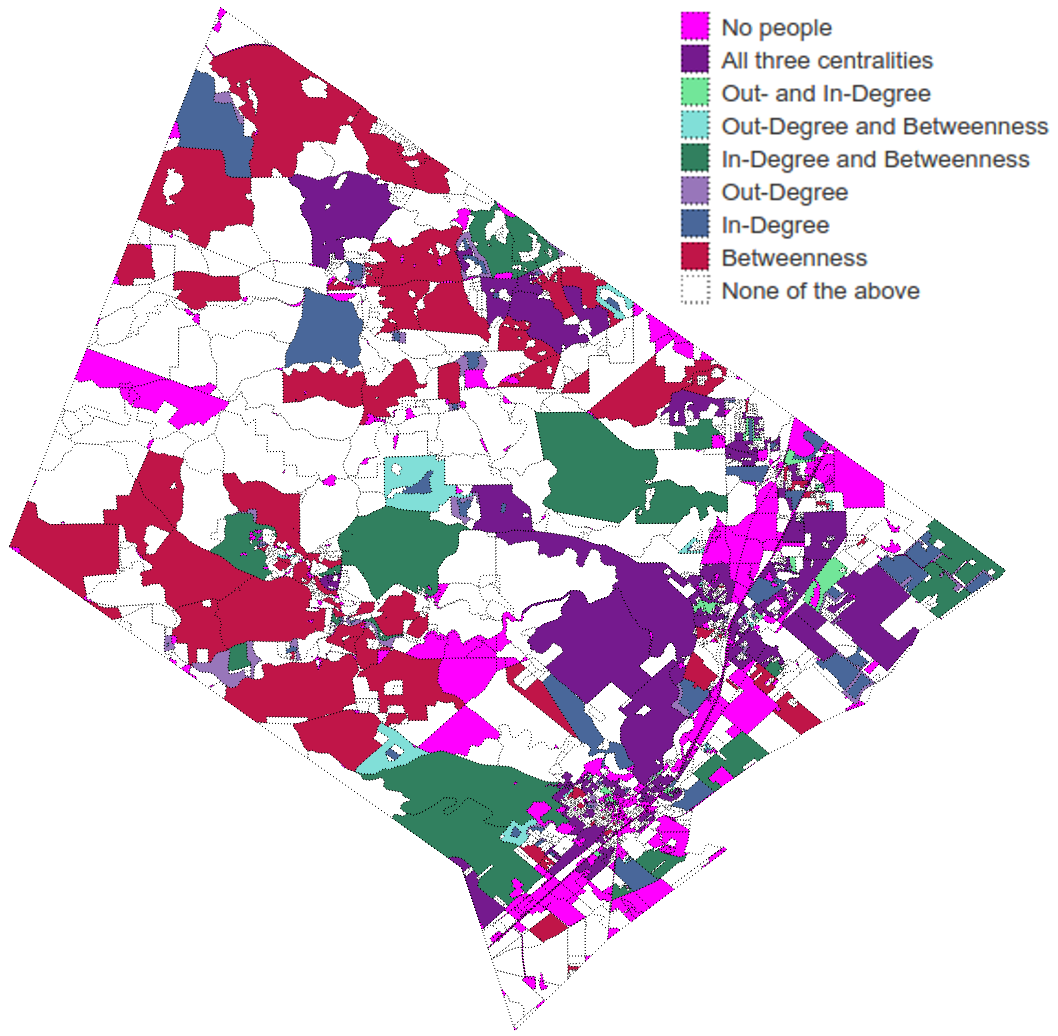}
  \caption{200 most central blocks in Hays County according to multiple centrality measures.}
  \label{fig:haysmap}
\end{figure}

We used maps generated by QuantumGIS, which highlighted approximately the top 10 percent most central blocks according to multiple centrality measures, to ensure that our model was producing realistic results.
These maps were also used as an aid in the analysis of our results.
A map of Hays County, showing the 200 most central blocks according out-degree, in-degree, and inverse betweenness centralities, is included as Figure \ref{fig:haysmap}.
This map was created by obtaining the list of the 200 most central census blocks under each centrality measure and marking them with distinct colors.
Blocks that were central according to multiple centrality measures received their own colors.
\input{tables-2}

\section{Discussion}
We found that in-degree and inverse betweenness centralities tended to be the most effective at containing epidemics.
At low vaccination levels, such as 30 percent and 50 percent, where almost all of the population becomes infected under all vaccination strategies, the peak infected percentage was used to compare the effectiveness of various strategies.
Lowering the number of infected people at one time reduces strain on public health infrastructure, allowing the cases that exist to be treated more effectively.
At higher vaccination levels, where no significant peak occurred, the state totals, along with the plots of the course of the epidemics, were used to determine effectiveness.
The peaks were not used to analyze vaccination effectiveness at the 75 percent and 90 percent intervention experiments, as the peaks were found to reflect the degree to which the disease had spread before the vaccines were distributed, and to have little relation to the effectiveness of allocation strategies.
These analyses allowed us to determine which of the tested vaccination strategies was most effective at containing the disease.

In all but one of our experiments, at least one of in-degree or inverse betweenness had lower peaks than the control, which was utilizing a random vaccination strategy.
Neither measure had a lower peak significantly more times than the other, which shows than inverse betweenness and in-degree centralities are both effective in various situations.
In-degree was significantly more effective in the intervention scenario, while inverse betweenness was more effective in the prevention scenario.
In-degree centrality allocated vaccines to those blocks that were the most vulnerable, that is, the most likely to be contacted.
Because it shielded vulnerable blocks from the disease, in-degree was more effective at containing epidemics that had already begun to spread throughout the county at the time of vaccination.
Inverse betweenness centrality, however, allocated vaccines to those blocks most likely to be involved in disease transmission.
This allocation strategy was more effective in the prevention scenario because it blocked off likely transmission paths, which would have allowed the epidemic to spread quickly.
Both of these targeting methods appear to be effective at reducing the severity of epidemics at the county level, though they are effective in different scenarios.
This conclusion agrees with that of Matrajt and Longini, who found that vaccinating the vulnerable and those likely to transmit the disease was an optimal strategy at the national level \cite{Matrajt2010}.

Out-degree centrality was consistently the least effective method of vaccination, with peaks higher than those of random vaccination in most of the 30 percent and 50 percent vaccination experiments.
This resulted from out-degree centrality's tendency to allocate vaccines to high-population areas that are likely to spread the disease once it reaches them, regardless of their probability of infection.
This result is supported by Figure \ref{fig:haysmap}, which shows out-degree targeting mainly small, urban blocks.
Figure \ref{fig:haysmap} also shows that out-degree centrality selected different blocks than the effective measures, which had many blocks in common.

At 75 percent vaccination, at least one of in-degree and inverse betweenness was more effective at halting disease spread than random vaccination in all experiments.
Similar to the 30 percent and 50 percent levels of vaccination, both measures were the most effective about half the time.
In-degree centrality tended to fall slightly faster than inverse betweenness at the beginning of the simulation, though inverse betweenness was more effective at stopping disease spread near the end of the simulation, a results that agrees with those of Matrajt and Longini \cite{Matrajt2010}.

Effectively no disease spread occurred under all measures, at 90 percent vaccination.
As at previous vaccination levels, in-degree and inverse betweenness were each most effective in half of the experiments.
The differences in effectiveness in the prevention and intervention scenarios were much less significant than at lower vaccination levels.
The lack of significant disease spread at 90 percent was a consequence of herd immunity, the primary factor influencing disease spread at high vaccination levels \cite{introepi}.
Due to the initial unvaccinated period, epidemics in the intervention scenario at 90 percent vaccination tended to last longer than those in the prevention scenario.

Eigenvector-based measures, especially eigenvector-in centrality, which targeted vulnerable blocks that were likely to spread the disease to other vulnerable blocks, showed an interesting trend: their effectiveness increased with county size and the prevalence of urban areas.
In fact, in the prevention scenario at 30 percent vaccination in Denton County, eigenvector-in centrality was the most effective, followed by inverse betweenness.
In Rockwall County, which is small and sparsely populated, these measures were the worst performers.
At 75 percent vaccination in Rockwall, eigenvector and eigenvector-in centralities produced epidemic curves similar to those observed at lower vaccination levels, which did not occur for the other measures.
In contrast, eigenvector-based measures, especially eigenvector-in centrality, were more effective in Hays and Denton counties, which are larger and more heavily urbanized.
In several experiments, eigenvector-in centrality outperformed random vaccination, though they were much less reliable than in-degree and inverse betweenness.
Eigenvector-in centrality's effectiveness increased in urbanized counties because eigenvector-based measures favor central blocks that are connected to other central blocks.
Clusters of vulnerable blocks, which eigenvector-in centrality selects, were prevalent in counties such as Hays and Denton than in Rockwall, leading to the increased effectiveness of that centrality measure.
However, as seen in Figure \ref{fig:rptotals}, eigenvector-based measures were still generally ineffective compared to other options.

In-degree centrality, at 30 percent and 50 percent vaccination, caused the longest delay in the peak of the epidemic.
This result arose because contacts within in-degree central blocks, which were more likely to occur than those within non-central ones, did not cause disease spread because those blocks were vaccinated.
This resulted in the disease taking longer to spread because it was forced into less probable paths.

In previous experiments with our simulator, we received unusual data, which we do not represent here.
Specifically, random vaccination was reported as the most effective method for both the intervention and prevention scenarios in Denton County. 
These results were most likely a result of the stochastic nature of our simulator. 
It is possible, however, that this result was caused by in-degree centrality's preference for the large number of highly-populated blocks in Denton County, which left large areas of the county unvaccinated.


Some of the limitations of our work are our assumptions that the population is homogenous within each block, that contact rates between census blocks are approximated by Equation \ref{eqn:weights}, and that the individual contact rate is constant.
Another limitation is that we are only testing one isolated county at a time.
These assumptions do not reflect reality, but are common approximations that have been used in computational models of epidemics such as the Global Stochastic Cellular Automata model \cite{Mikler2005}.

Future work will primarily focus on addressing the limitations described above.
Methods for simulating non-homoge-\ \ nous populations \cite{Reyes-Silveyra2011,Indrakanti2012} will be incorporated into our model.
Alternative centrality measures and more accurate weighting formulas will be investigated.
We will also attempt to determine what methods would produce results that surpass random vaccination in cases that are currently outliers.
\bibliographystyle{acm}
\bibliography{Vaccine_distribution,Centrality_Measures,Epidemiological_applications_of_centrality,Computational_epidemiology,sources}
\end{document}

%% file: tables-2.tex
\begin{table*}
  \centering
  \scalebox{0.9}{%
  \begin{tabular}{|l |r r r r r|}
    \hline
    & \multicolumn{5}{|c|}{Prevention Scenario: Rockwall County} \\ \hline
    \% pop (\# vaccines) & No vaccination & 30\% (23501) & 50\% (39168) & 75\% (58752) & 90\% (70503) \\ \hline
    Out-Degree & \multirow{6}{*}{59.5177 (0.5823)} & 31.8124 (0.3003) & 12.9924 (0.2802) & 0.0207 (0.0079) & 0.0073 (0.0030)\\
    Inverse Betweenness & & 29.2634 (0.3717) & \textbf{9.6473} (0.2685) & \textbf{0.0181} (0.0086) & \textbf{0.0051} (0.0051)\\
    Eigenvector & & 32.5062 (0.3980) & 13.9375 (0.3912) & 1.4525 (0.7278) & 0.0079 (0.0038)\\
    Eigenvector-In & & 29.9352 (0.4541) & 12.5988 (0.3685) & 1.1630 (0.6695) & 0.0077 (0.0037)\\
    In-Degree & & \textbf{29.1990} (0.2862) & 10.5134 (0.2083) & 0.0220 (0.0106) & 0.0068 (0.0030)\\
    Random & & 29.7944 (0.4934) & 10.5293 (0.2976) & 0.0184 (0.0135) & 0.0060 (0.0081)\\
    \hline & \multicolumn{5}{|c|}{Hays County} \\ \hline
    \% pop (\# vaccines) & No vaccination & 30\% (47132) & 50\% (78553) & 75\% (117830) & 90\% (141396)\\ \hline
    Out-Degree & \multirow{6}{*}{56.9851 (1.1604)} & 32.6743 (0.3588) & 13.8937 (0.1090) & 0.0208 (0.0066) & 0.0053 (0.0029)\\
    Inverse Betweenness & & \textbf{26.1614} (0.5565) & \textbf{9.3640} (0.1403) & \textbf{0.0172} (0.0071) & \textbf{0.0059} (0.0036)\\
    Eigenvector & & 28.1819 (0.2369) & 11.1222 (0.1744) & 0.0183 (0.0058) & 0.0052 (0.0023)\\
    Eigenvector-In & & 27.7457 (0.3377) & 10.8385 (0.1924) & 0.0190 (0.0061) & 0.0050 (0.0025)\\
    In-Degree & & 28.0340 (0.4022) & 9.9216 (0.1894) & 0.0175 (0.0057) & 0.0062 (0.0025)\\
    Random & & 29.3217 (0.3645) & 10.5042 (0.2072) & 0.0244 (0.0115) & 0.0095 (0.0094)\\
    \hline & \multicolumn{5}{|c|}{Denton County}\\ \hline
    \% pop (\# vaccines) & No vaccination & 30\% (198784) & 50\% (331307) & 75\% (496960) & 90\% (596353)\\ \hline
    Out-Degree & \multirow{6}{*}{62.3313 (0.4791)} & 34.9600 (0.1829) & 14.4588 (0.0759) & 0.0196 (0.0028) & 0.0059 (0.0014)\\
    Inverse Betweenness & & 32.2106 (0.1637) & 13.1199 (0.0911) & \textbf{0.0178} (0.0040) & \textbf{0.0057} (0.0020)\\
    Eigenvector & & 34.4785 (0.1946) & 15.3266 (0.0855) & 0.0374 (0.0359) & 0.0058 (0.0014)\\
    Eigenvector-In & & \textbf{32.1597} (0.1371) & 13.8856 (0.1084) & 0.0212 (0.0039) & 0.0058 (0.0014)\\
    In-Degree & & 32.3480 (0.1617) & \textbf{12.8719} (0.1070) & 0.0190 (0.0024) & \textbf{0.0057} (0.0016)\\
    Random & & 32.3165 (0.2183) & 13.0035 (0.0921) & 0.0186 (0.0086) & 0.0061 (0.0038)\\
    \hline
  \end{tabular}%
  }
  \caption{For the prevention scenario, average peak infected percentage with standard deviation in parentheses. Lowest values in boldface. Because there were no vaccines to be distributed, there is only one value for no vaccination.}
  \label{tab:peaksp}
\end{table*}

\begin{table*}
  \centering
  \scalebox{0.9}{%
  \begin{tabular}{|l |r r r r r|}
    \hline
    & \multicolumn{5}{|c|}{Intervention Scenario: Rockwall County} \\ \hline
    \% pop (\# vaccines) & No vaccination & 30\% (23501) & 50\% (39168) & 75\% (58752) & 90\% (70503)\\ \hline
    Out-Degree & \multirow{6}{*}{59.3418 (1.0607)} & 32.0758 (0.5265) & 13.5754 (0.4445) & 1.7544 (0.5848) & 1.6049 (0.5464)\\
    Inverse Betweenness & & 29.4420 (0.5808) & \textbf{10.1374} (0.3663) & 1.8280 (0.6144) & 1.6352 (0.5571)\\
    Eigenvector & & 32.8470 (0.7541) & 15.2428 (0.4611) & 2.0908 (0.3910) & 1.6432 (0.5579)\\
    Eigenvector-In & & 30.7960 (0.7225) & 13.8876 (0.4673) & 1.8606 (0.4803) & 1.6393 (0.5607)\\
    In-Degree & & \textbf{29.3326} (0.7046) & 10.6855 (0.4030) & \textbf{1.7150} (0.5744) & \textbf{1.6143} (0.5672)\\
    Random & & 30.5434 (0.7600) & 11.9360 (0.3728) & 1.7970 (0.6138) & 1.6320 (0.5323)\\

    \hline & \multicolumn{5}{|c|}{Hays County} \\ \hline
    \% pop (\# vaccines) & No vaccination & 30\% (47132) & 50\% (78553) & 75\% (117830) & 90\% (141396)\\ \hline
    Out-Degree & \multirow{6}{*}{58.0324 (1.8390)} & 32.3683 (0.7623) & 13.4622 (0.3566) & 1.8578 (0.6219) & 1.7342 (0.5743)\\
    Inverse Betweenness & & \textbf{27.5758} (0.9813) & 10.2484 (0.4134) & 1.9364 (0.6338) & 1.7479 (0.5707)\\
    Eigenvector & & 28.9617 (0.8530) & 12.1266 (0.4515) & 1.8978 (0.6402) & 1.7367 (0.5627)\\
    Eigenvector-In & & 28.4758 (0.7967) & 11.9360 (0.4113) & 1.8689 (0.6336) & 1.7339 (0.5832)\\
    In-Degree & & 28.2601 (0.9398) & \textbf{9.8329} (0.4272) & \textbf{1.8171} (0.6087) & \textbf{1.7230} (0.5756)\\
    Random & & 29.0436 (1.0581) & 11.3709 (0.6273) & 1.9226 (0.6457) & 1.7459 (0.5652)\\

    \hline & \multicolumn{5}{|c|}{Denton County}\\ \hline
    \% pop (\# vaccines) & No vaccination & 30\% (198784) & 50\% (331307) & 75\% (496960) & 90\% (596353)\\ \hline
    Out-Degree & \multirow{6}{*}{62.8987 (0.5339)} & 34.8273 (0.4330) & 14.5756 (0.1599) & 1.7676 (0.3085) & 1.6589 (0.2921)\\
    Inverse Betweenness & & 32.8160 (0.2959) & 13.5322 (0.1470) & 1.8390 (0.3245) & 1.6654 (0.3033)\\
    Eigenvector & & 34.6035 (0.2898) & 15.4285 (0.1743) & 1.7687 (0.3033) & 1.6604 (0.2939)\\
    Eigenvector-In & & 32.3856 (0.2001) & 14.1328 (0.1577) & 1.7384 (0.3112) & 1.6528 (0.2914)\\
    In-Degree & & \textbf{32.2367} (0.2579) & \textbf{13.0264} (0.1090) & \textbf{1.7361} (0.3060) & \textbf{1.6519} (0.2921)\\
    Random & & 33.0462 (0.4129) & 13.6822 (0.1478) & 1.8266 (0.3223) & 1.6691 (0.3046)\\
    \hline
  \end{tabular}%
  }
  \caption{For the intervention scenario, average peak infected percentage with standard deviation in parentheses. Lowest values in boldface. Because there were no vaccines to be distributed, there is only one value for no vaccination.}
  \label{tab:peaksi}
\end{table*}